# Observation of wave amplification and temporal topological state in a genuine photonic time crystal

Jiang Xiong[1], Xudong Zhang[1,5], Longji Duan[1], Jiarui Wang[1], Yang Long[2], Haonan Hou[1], Letian Yu[2], Linyang Zou[3,*], Baile Zhang[2,4,†]

[1]School of Physics, University of Electronic Science and Technology of China, Chengdu 611731, China.

[2]Division of Physics and Applied Physics, School of Physical and Mathematical Sciences, Nanyang Technological University, Singapore 637371, Singapore.

[3]School of Electrical and Electronic Engineering, Nanyang Technological University, Singapore 639798, Singapore.

[4]Centre for Disruptive Photonic Technologies, Nanyang Technological University, Singapore, 637371, Singapore.

[5]School of Systems Science and Engineering, Sun Yat-sen University, Guangzhou 510275, China

*linyang001@e.ntu.edu.sg
†blzhang@ntu.edu.sg

**Photonic time crystals (PTCs) are materials whose dielectric permittivity is periodically modulated in time, giving rise to bandgaps not in energy--as in conventional photonic crystals--but in momentum, known as *k*-gaps. These *k*-gaps enable wave amplification by extracting energy from temporal modulation, offering a mechanism for coherent light generation that bypasses traditional optical gain. PTCs also extend the concept of topological insulators to the time domain, inducing a temporal topological state at the mid-gap of the *k*-gap, characterized by the Zak phase—a topological invariant originally defined for spatial lattices. Here, we experimentally demonstrate the properties of a *k* gap in a genuine PTC, realized in a dynamically modulated transmission-line metamaterial. Wave amplification within the *k*-gap is observed, with an initial power spectrum narrowing and shifting toward the gap. To probe the mid-gap**

topological state, we introduce a temporal interface separating two PTCs with distinct topological phases. The measured phase shift between time-reflected and time-refracted waves, together with the temporal confinement of the topological state, provides direct evidence of nontrivial temporal topology. By integrating k-gap amplification with time-domain topological features, our work opens new avenues for light generation and manipulation in time-varying photonic materials.

# I. Introduction

Crystals generally refer to structures with periodicity in space. In photonics, it has been known since the 1980s that photonic structures with dielectric permittivity periodically modulated in space, commonly referred to as photonic crystals [1–3], can give rise to photonic bandgaps—frequency ranges where waves cannot propagate but only decay evanescently—similar to the energy bandgaps in semiconductors [4] (Fig. 1a). Recently, inspired by topological insulators [5], photonic bandgaps have been endowed with topological properties, leading to the development of photonic topological insulators [6–8]. As a defining feature, the boundary between two topologically distinct photonic topological insulators hosts topologically protected edge states (Fig. 1b), as determined by the bulk-boundary correspondence, a fundamental principle in topological physics. Taking the simplest one-dimensional (1D) photonic topological insulator [9] as an example, the existence of an in-gap topological edge state introduces a peak inside the topological bandgap in the wave transmission spectrum through such a 1D system (Fig. 1c).

Recently, time has been proposed as an additional dimension of crystallization in photonic time crystals (PTCs) [10–17], which are photonic materials whose permittivity varies periodically in time. It has been shown that PTCs can exhibit bandgaps in momentum, known as $k$-gaps (see Fig. 1d). In contrast to spatially evanescent modes in conventional photonic bandgaps, these $k$-gaps can accommodate temporally amplifying modes. Unlike optical parametric amplifiers, which requires precise phase matching [18], such amplification is non-resonant over the entire $k$ gap, but is maximized at the mid-gap, providing a mechanism of coherent light generation [19–21] (e.g., lasers) by drawing energy from time modulation rather than gain media.

Band topology can also be extended to $k$-gaps in PTCs, giving rise to topological phenomena in the time domain. It has been proposed [22] that a $k$-gap carries band topology characterized by the Zak phase [23], a topological invariant used to describe 1D topological insulators. In this context, a nontrivial Zak phase enables a PTC to function as a "temporal" topological insulator. At a temporal interface between two PTCs with distinct Zak phases, a mid-gap temporal edge state can arise in the $k$-gap (Fig. 1e), analogous to the topological edge state in spatial topological insulators [7]. This temporal topological edge state confines energy in time rather than in space and extends the principle of bulk-boundary correspondence to the time domain. The existence of this in-gap topological state leads to a dip in the transmission spectrum through the 1D system (Fig. 1f).

Here, we present an experimental study of the *k*-gap properties in a PTC realized in a dynamically modulated transmission-line metamaterial [24–26]. We employ an advanced differential modulation strategy that allows for the direct observation of the space-time evolution of wave dynamics in the PTC. Through space-time Fourier transforms, we capture the temporal band structure, revealing the *k*-gaps. We demonstrate wave amplification within the *k*-gap, where an arbitrary initial power spectrum narrows and shifts toward the *k*-gap center, confirming the mechanism of coherent light generation via temporal modulation. We then experimentally probe the temporal topology of the *k*-gap, characterized by the Zak phase, from two perspectives. First, similar to previous studies on 1D photonic topological insulators [9], the time-reflected and time-refracted waves through the PTC experience different phase shifts, providing clear evidence of the quantized Zak phase. Second, we introduce a temporal interface separating two PTCs with distinct Zak phases. A clear topological state that confines energy in time and induces a dip in the transmission spectrum is observed experimentally. This marks the first experimental observation of a temporal topological edge state in a material platform.

## II. The PTC Realization and Experimental Setup

In this work, we employ a microstrip dynamic transmission line (DTL) to realize a PTC. Fig. 2a illustrates the DTL geometry. A metallic microstrip trace line is photoetched onto an FR4 substrate with a dielectric constant of $\varepsilon_r = 2.2$, and is periodically loaded with 64 pairs of shunt varactors and series inductors. The varactors provide non-linear, bias-voltage-dependent capacitance, while the inductors are placed to extend the optical path within each unit cell. Under the long-wavelength approximation, the DTL behaves as an effective uniform media under DC bias [24,27,28]. In Fig. 2b, We compute the effective permittivity using the field-circuit equivalence [29] across various DC bias levels. In this experiment, we modulate the DTL within a quasi-linear range (indicated by the blue line) to achieve a sinusoidal varying permittivity modulation.

Unlike conventional DTL designs, we adopt a differential modulation scheme in which each pair of varactors is reverse-biased with opposite polarities and symmetrically placed on either side of the microstrip trace line in each unit cell. Fig. 2c depicts the equivalent circuit model for several DTL unit cells. $L_0, C_0$ represent intrinsic unit-length series inductance and shunt capacitance of the host microstrip line. $L$ denotes the loaded lumped inductor. The blue and orange color varactors have the opposite polarities and $C(t)$ represents the time-varying capacitance of the varactor diodes. Each varactor is shunted to the ground via a band-stop filter (BSF) resonant at the modulation frequency, thereby isolating the modulation signal. This configuration allows the modulation contributions from each varactor to cancel each other out and leave a clean environment for dynamic signal detection without diminishing the modulation depth.

To achieve precise temporal control over the effective permittivity in the time domain, as shown in Fig. 2d, we use an arbitrary waveform generator (AWG) to produce both the probe and modulation signals. A three-stage power-dividing network is designed to provide coherent modulation signal $V_A(t)$ and $V_B(t)$ for each pair of varactors. Fig. 2e shows the fabricated DTL, with an inset providing a magnified view of a unit cell. For experimental measurement, we integrate a field scanner, the DTL sample, and the power dividing network into a compact scaffold-like measurement platform (Fig. 2f). This differential modulation scheme, synchronized via an externally generated clock signal, enables direct observation of the genuine spatial-temporal evolution of electromagnetic waves within the time-modulated DTL. Further details on the effective permittivity calculation, experimental setup, and the differential modulation performance, are provided in the supplementary materials.

### III. Experimental Results

We first demonstrate the dispersion of a static DTL, Fig. 3a presents the experimentally extracted band structure of the static DTL, with all varactors biased at a DC voltage of 0.6V. A broadband probe pulse is initially injected into the DTL, and we obtain the static dispersion $v(\overline{\omega}, \overline{k})$ by a 2D Fourier transform of the recorded spatiotemporal voltage $v(x_l, t)$ along the meandered line, where $x_l$ denotes the position index. Throughout this work, frequencies are normalized by the modulation frequency $\Omega$, and the wavevector $k$ is accordingly scaled by the unit wavevector $k_0 = \sqrt{\varepsilon_{eff}}\Omega/c_0$. A frequency bandgap emerges at $\overline{\omega} = 1$ as a consequence of the coupling between the transmission line and the periodically loaded BSFs. By modeling the BSFs as effective LC resonators, we numerically calculate the dispersion relation [29,30] (solid line in Fig. 3a), which shows good agreement with the measurement. At $\overline{\omega} = 0.5$, the dispersion remains linear, ensuring the *k*-gap formation in the subsequent modulation.

When the DTL is sinusoidally modulated in time, a PTC and its Floquet bands emerge, with *k* gaps open in momentum space. To fully excite the Floquet eigenmode in positive and negative momentum space, we inject a broadband Gaussian pulse train from both ports of the DTL simultaneously. By converting the measured space-time signal into the frequency-momentum domain, we obtain the Floquet band structure shown in Fig. 3b. The measured dispersion reveals two hallmark features of a PTC: (1) periodic replication of bands along the frequency axis and (2) opening of momentum band gaps at $\bar{\omega}$ = 1/2. Using the temporal Transfer Matrix Method (TMM) [31] and Floquet theory [32], the theoretically calculated dispersion relation (yellow solid curves in Fig. 3b) exhibits excellent agreement with the experimental data. A horizontal stripe at $\bar{\omega}$ = 1 is attributed to minor leakage of the modulation signal, which is not fully canceled by the differential modulation scheme. The details of the theoretical calculation are provided in Supplementary Material.

The *k* gap modes, by virtue of their non-resonant amplification mechanism, can spontaneously generate coherent standing waves regardless of the input [19]. When a wave with abundant momenta is incident into a PTC, the momenta in the band region can couple to propagating-wave eigenmodes without experiencing amplification, whereas the momenta within the *k* gaps induce exponential amplification, giving rise to standing waves [28,33] and spontaneous formation of spatiotemporal pattern [22]. To demonstrate these unique phenomena, we excite the DTL with a quasi-monochromatic wave whose central momenta lies well outside the *k* gap. Due to the finite spatial size of DTL, a small fraction of the wave's spatial spectrum leaks into the *k* gap region. We record the full spatial-temporal evolution of the wave in DTL and visualize the dynamics in Fig. 3c, with particular moments of four initial times plotted in Fig. 3d. Initially, before the modulation starts, the incident wave propagates freely as a traveling wave ($t = -10T$). Upon activation of the modulation, amplification commences, and a standing-wave pattern emerges ($t = 120T$), accompanied by a reduction in the effective wavelength, validating a dominant momentum conversion. For further analyzing the dynamics, we extract the temporal evolution of the normalized spatial power spectrum by applying a spatial Fourier transform at each time step. As shown in Fig. 3e, although the input momenta ($\bar{k}_{gap} = 0.41$) lies far away from the *k* gap ($\bar{k}_{gap} = 0.57$), the *k* gap momenta dominates the power spectrum shortly after the modulation starts, overshadowing all out-of-gap components. This observation confirms the spontaneous emission behavior of PTCs as predicted theoretically [19]. Additionally, we plot the temporal evolution of the spectral power for both the incident and mid-gap momentum components in log scale (Fig. 3f) and fit them with a straight line, demonstrating the predicted exponential amplification of the *k*-gap mode. Note that a discrepancy between the measured result and the perfect exponential growing dashed line becomes apparent after a sufficiently long time (marked in orange in Fig. 3f). We attribute such a saturation to practical limitations in the experimental setup: 1. Finite modulation energy input from power amplifiers. 2. Intrinsic loss in multiple components (DC supply, BSF, etc.). 3. Non-linear capacitance-voltage response at large amplitude. Throughout the measurement, we carefully design the modulation scheme so that all the expected physical events happen long before such saturation emerges.

Interestingly, PTCs also exhibit band topology in the temporal Brillouin zone, characterized by the Zak phase. Firstly, as theoretically established in both PTCs [22] and spatial photonic crystals [9], in a 1D periodic system with lattice mirror symmetry, the Zak phase of each band is quantized to 0 or $\pi$. Furthermore, the sign of a wave scattering observable, i.e. the reflection wave phase for a spatial photonic crystal [9] or the phase difference between the time-reflected and time-refracted waves for a PTC [22], can be used to determine the Zak phase of bulk bands. In the context of PTCs, this relation is explicitly given by:

$$sgn(\varphi_n) = (-1)^{n+l} exp(i \sum_{m=1}^{n-1} \theta_m^{Zak})$$

where $\varphi_n$ is the phase difference between the time-reflected and time-refracted waves for the $n^{th}$ *k* gap (with the lowest gap number index as 1), evaluated at the last temporal interface between a PTC and the subsequent static medium. *l* accounts for the number of band crossings under the $n^{th}$ momentum gap ($l = 0$ in our study), and $\theta_m^{Zak}$ represents the Zak phase of $m^{th}$ band. When two PTCs with distinct Zak phases are temporally cascaded, a mid-gap topological edge state emerges inside the *k* gap. In our experiment, the sinusoidal modulation opens only the first momentum bandgap. Therefore, we extracted $\varphi_1$ (referred to simply as $\varphi$ hereafter) to characterize the *k*-gap topology. In the experiment, due to the finite spatial length of the DTL, the momentum resolution inside the *k* gap is limited, making it difficult to resolve the intensity/phase spectrum with high precision. To circumvent this issue, we repeat the measurement at different modulation frequencies and effectively increase the number of measurable points in momentum space. Detailed methodology and the phase extraction procedure are described in the Supplementary Materials. The experimentally extracted values of $\varphi$ at nine momenta points in the $k$ gap are shown in Fig. 4a with blue triangles and red diamonds corresponding to two PTC configurations with $\theta_1^{Zak} = 0$ and $\pi$, respectively, referred to as PTC1 and PTC2 with opposite temporal lattice mirror symmetry center. The theoretical relation of $\varphi$ and $k$ calculated by Transfer Matrix Method, are plotted with blue and red solid lines for comparison. The experimental results agree well with the theory. As expected, PTC1 and PTC2 exhibit the same monotonicity in $\varphi(k)$ but with opposite signs, highlighting their distinct $k$-gap topologies.

Finally, we observe the emergence of a temporal edge state at the interface between PTC1 and PTC2. In the context of spatial photonic crystals, it has been demonstrated [9] that the appearance of an unimodular transmission spike at a specific frequency for two photonic crystals with distinct gap topologies ensures a topological edge state confined in space, which is also the point where the summation of the reflection phases of the respective spatial photonic crystals approaches zero. By analogy, a similar phenomenon is expected in the temporal domain: the zero point of the summation of $\varphi$ for two PTCs with distinct Zak phase (indicated by the black line and dots in Fig. 4a) predicts a critical momenta $k_c$, where a temporal topological edge state arises. Unlike that in SPCs, the temporal topological edge state in PTCs manifests as a dip in the total energy spectrum, while the remaining momentum components within the $k$ gap exhibit high transmission. Fig. 4b presents both the theoretical (pink solid line) and experimentally extracted (red dot) total energy spectrum. The rationale for adopting total energy |T+R| instead of transmission |T| is explained in Supplementary Materials. As expected, a dip appears at $\bar{k} = 0.564$ in the total energy spectrum, which is close to the theoretical prediction $k_c$ (marked by dashed green lines in both Figs. 4a and b). To probe the associated field dynamics, we excite the DTL at this critical momentum and plot the spatiotemporal field evolution in Fig. 4c. The field shows temporal localization centered at $t = 0$, consistent with a temporally localized edge state. We extract the temporal power evolutions of the edge state momenta and plot it in Fig. 4d (blue curve). Prior to the temporal interface between PTC1 and PTC2, the energy exhibits exponential growth as a $k$-gap mode. After the temporal switching, the energy undergoes exponential decay, forming a temporal localization profile. This distinct growth-decay pattern constitutes the first direct observation of a temporal topological edge state in a genuine PTC as a material platform. Long after the temporal switching, the energy is finally re-amplified, which originates from the simultaneous excitation of temporal growth and decaying $k$- gap eigenmodes. Note that the temporal asymmetrical growth-decay profile originates from intrinsic losses in the DTL, which suppresses the pre-interface amplification rate and enhances the post-interface decay rate. For comparison, the temporal power evolution of an extended bulk state at $\overline{k} = 0.19$ is also shown as the red curve in Fig. 4d, rendering a field temporal pattern with a temporally stable magnitude. Field evolutions of several other gap momentum components are also examined and are provided in the supplementary materials. Compared to the pronounced temporal localization observed in Fig. 4d, these components exhibit significantly shallower temporal peaks and correspondingly higher transmission, consistent with their weaker confinement in time.

## IV. Conclusion

The above results conclusively demonstrate two defining properties of a PTC: *k*-gap amplification and temporal topology, both observed in a genuine PTC implemented via a dynamically modulated transmission-line metamaterial platform. The *k*-gap amplification reveals a fundamentally distinct mechanism for coherent light generation that circumvents the need for conventional gain media. The temporal topology extends the concept of topological insulators into the time domain, effectively realizing a "temporal" topological insulator. Together, these findings lay the groundwork for future investigations into light–matter interactions in time-varying photonic materials [35,36]—including potential applications in lasing [19,37]—and offer new insights into the physics of space–time topological phases [38–41].

## Acknowledgment

The authors would like to thank Prof. Hao Hu, and Lidong Huang for the helpful discussions, and Chengdu YUEXIANG Technology Co., Ltd and Chengdu 6914 Technology Co., Ltd for their support in the experiment assembly and technical support in the experiment. Y.L. gratefully acknowledges the support of the Eric and Wendy Schmidt AI in Science Postdoctoral Fellowship, a Schmidt Futures program.

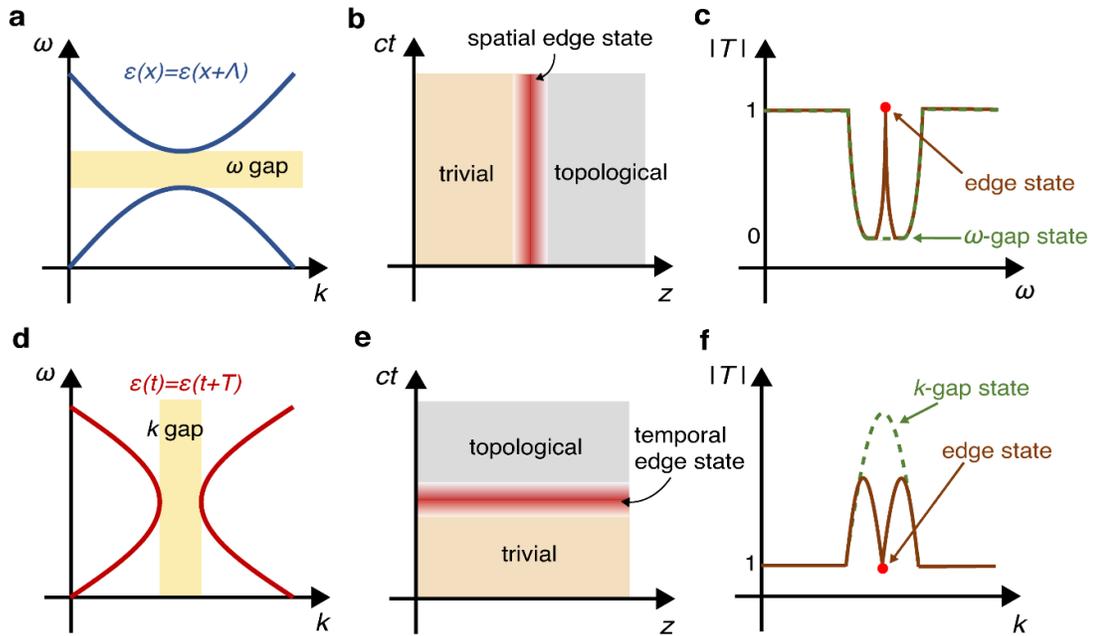

**Fig.1| Comparison between a conventional (spatial) photonic crystal and a photonic time crystal. a,** Energy bandgap in a spatial photonic crystal. **b,** Spatial topological edge state in a 1D spatial topological domain wall. The energy is localized in space while extended in time domain. **c,** (brown) Transmission spectrum of a spatial topological domain wall in (b), where the transmission peak in the bandgap corresponds to the topological edge state. (green) Transmission spectrum of a pure spatial photonic crystal, in the whole bandgap frequency range most of the energy is reflected back. **d,** $k$-gap in a photonic time crystal. **e,** Temporal topological edge state in a temporal topological domain wall structure. The total energy is localized in time while extended in space. **f,** (brown) Transmission spectrum of the temporal topological domain-wall structure in (e) shows a pronounced dip to zero at the mid-gap edge state. (green) Transmission spectrum of a single photonic time crystal, where the transmitted energy is amplified in the whole $k$ gap.

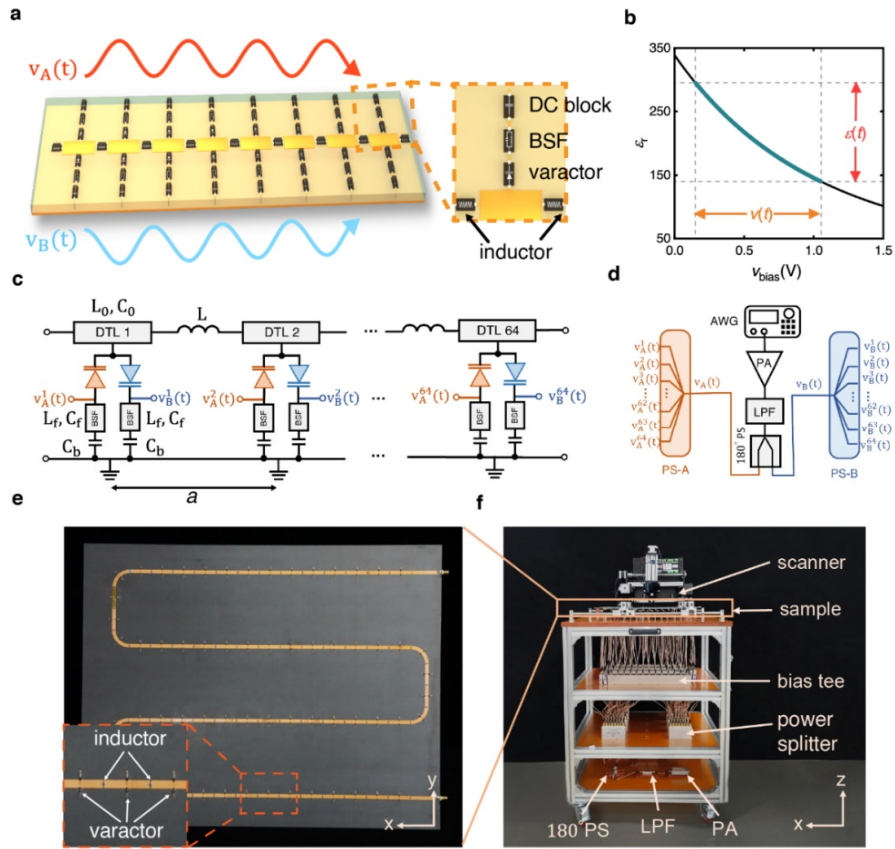

**Fig.2| A genuine photonic time crystal implemented in a microwave transmission-line metamaterial. a,** Schematic view of the dynamic transmission line. It is a microstrip line periodically loaded with shunt hyper abrupt union varactors and series inductors. The microstrip line applied has a 4.9mm wide trace line photoetched on a 1.575mm thick FR4 substrate with dielectric constant $\varepsilon_r$=2.2 and loss tangent tan$\delta$=0.001 **b,** $\varepsilon_r$ of the effective medium in **c** at $\omega = 2\pi \times 65MHz$ with the varactor bias voltage. The variation spans of a sinusoidal time-varying $v(t)$ and the resultant quasi-harmonic $\varepsilon_r(t)$ are denoted by orange and red arrows, respectively, and they are both within the relatively linear portion of the curve. **c,** Equivalent circuit model for the dynamic transmission-line unit cell and the resultant effective time-varying medium. The upper portion shows the equivalent circuit model with symmetric topology for the transmission-line unit cell in **a**. **d,** Schematic of three-stage power-dividing network, where the modulation signal is first spitted by a 180-degree phase shifting power divider, and enters two stage 8-way power dividers, finally enter the bias tee for varactor feeding. **e,** Fabricated sample of the meander DTL comprising 64-unit cells. A close-up view of the unit cell is shown in the inset. **f,** Scaffold-like measurement platform, showing the 3-stage power dividing network in d. A scanner synced with AWG and an oscilloscope is set above the sample for space-time scanning in situ measurement.

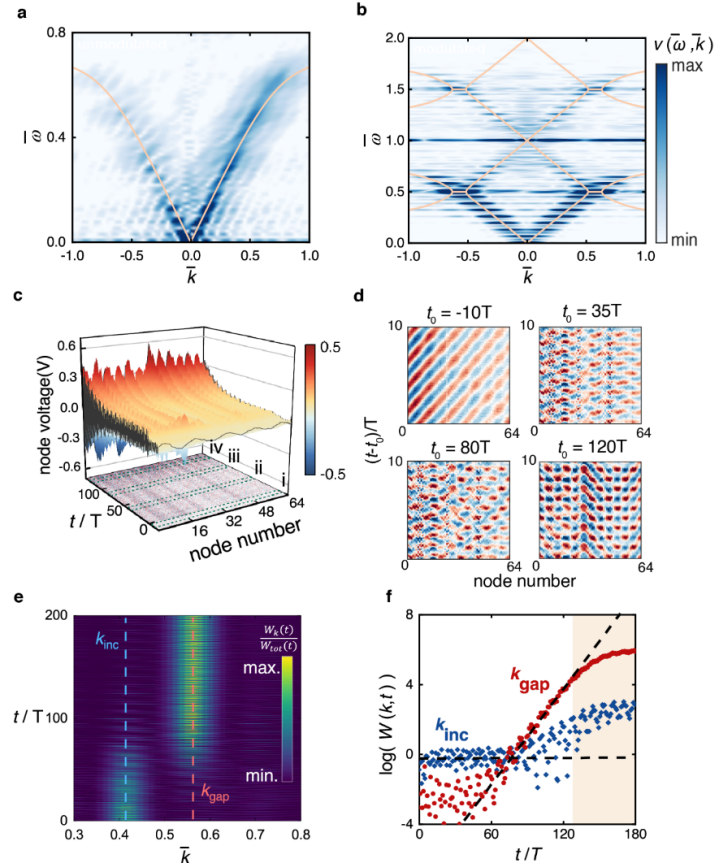

**Fig.3| Observation of *k*-gap amplification. a,** Static dispersion of the microstrip line. The heatmap is the normalized band structure of the transmission line with DC bias only, which is obtained via a Fourier transform of the spatiotemporally measured voltage along the line. The frequency is normalized by $\Omega$, and the wave number is accordingly scaled by that of a wave propagating in the static line without BSF loading at this frequency. Theoretically calculated dispersion is overlaid on the measured results. **b,** Band structures of the DTL when the temporal modulation is added. The heatmap shows the measured normalized band structure, and the theoretical band structures is overlaid as the solid yellow line for comparison. *k* gaps open at $\bar{k} = \pm 0.57$. **c,** Spatiotemporal evolution of the node voltage when excited at *k* gaps. **d,** Four zoomed in space-time sections in **c**. The wave in *k* gaps exhibits a mode transition from propagating mode to a standing wave profile. **e,** Temporal evolution of power spectrum along the DTL when an initial signal of wide spectrum is used for excitation. The momenta of the input signal and the experimentally measured *k*-gap center are denoted by blue and red dashed lines, respectively. **f,** Experimentally extracted *k*-gap and excitation momenta energy evolution in a log scale. The *k*-gap mode undergoes exponential growing while the excitation momenta energy remains uniform in time. The orange box indicates the saturation region where the energy stops growing.

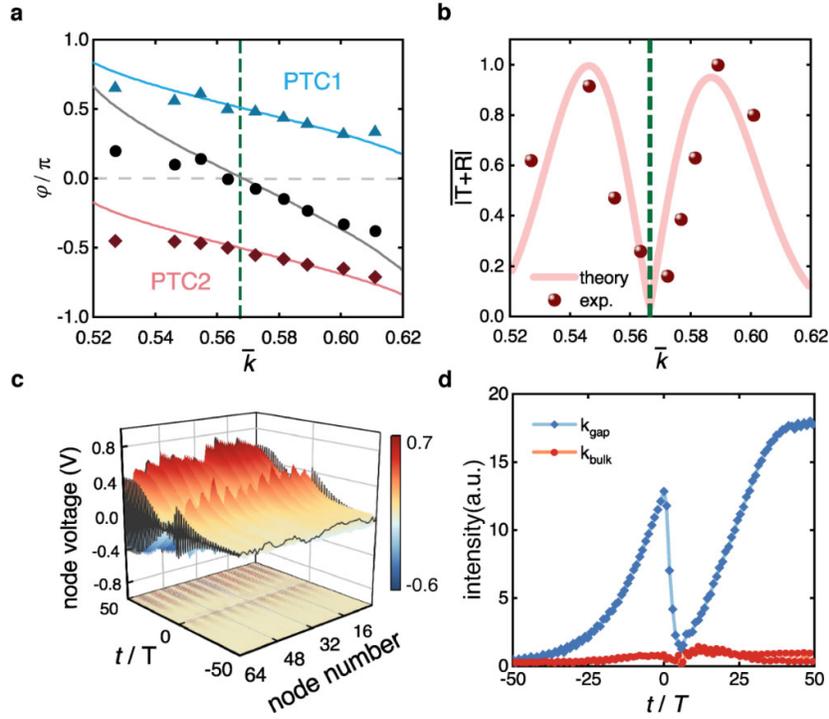

**Fig.4| Observation of the temporal topological edge state. a,** spectrum of the phase difference between the time-transmitted and time-reflected waves inside the *k* gap. Two photonic time crystals with identical modulation parameters but an opposite initial modulation phase, denoted as PTC1 and PTC2, are investigated. Experimentally extraction of such a phase difference for PTC1 and PTC2 are plotted with blue triangle and red diamonds. Corresponding theoretically calculated results are plotted with blue and red solid lines for comparison. Phase differences in the two PTCs exhibit the same monotonicity but opposite signs, indicating distinct topologies. The summation of the two phase differences is plotted with black dots and a solid line, respectively. **b,** Experimentally extracted (red dot) and theoretically calculated (red line) normalized total energy spectrum inside the *k* gap. A topological edge state appears at the dip in the spectrum, exhibiting unimodular transmission, whose momenta is noted with a green dashed line, same as the green dashed line in **a**. **c,** Measured space-time field evolution in temporally cascaded photonic time crystals. PTC1 is converted to PTC2 at $t = 0$. **d,** Temporal evolution of the edge state with $\overline{k} = 0.5633$ inside the *k*-gap (blue line) and the bulk state with $\overline{k} = 0.19$ (red line) in the cascading system in **c**. A localized temporal edge state emerges at t=0 in *k*-gap, while an extended state is obtained in bulk band.